\documentclass[10pt,twocolumn,letterpaper]{article}
\usepackage[T1]{fontenc}
\usepackage{lmodern}
\usepackage[margin=0.75in,columnsep=0.25in]{geometry}
\usepackage{graphicx}
\usepackage{amsmath,amssymb}
\usepackage{cite}
\usepackage{microtype}
\usepackage{url}
\graphicspath{{images/}}
\date{}

\begin{document}

\title{Energy-Efficient Geographic Multicast Routing in Cognitive Wireless Sensor Networks}

\author{Farzan Heidari\\ {\small Independent Researcher}
\and Farzaneh Heidari\\ {\small Mila, Universit\'e de Montr\'eal, Montr\'eal, Canada}}

\maketitle

\begin{abstract}
Energy consumption is a major design concern in wireless sensor networks. Cognitive radio provides opportunistic access to licensed spectrum, while multicast routing allows several destinations to share forwarding paths. In this paper, we propose two geographic multicast routing protocols for cognitive wireless sensor networks, CCast and PCast. CCast combines multicast-tree construction with a latency-oriented spectrum-selection metric. PCast uses a normalized link-power cost to select routes and can avoid regions where primary users occupy lower-frequency bands. Both protocols use local neighbor information and destination coordinates, without requiring a complete network-topology map. Simulations compare the proposed protocols with geographic multicast routing (GMP) and multiple unicast routing. In simulations, CCast and PCast reduce energy consumption by 21\% and 26\% relative to GMP, at the cost of longer per-destination routes and higher delivery delay.
\end{abstract}
\medskip
\noindent\textbf{Keywords}\quad Multicast routing, cognitive radio, wireless sensor networks.

\section{Introduction}

Wireless sensor networks (WSNs) support a wide variety
of applications, such as environment monitoring, smart battlefield,
health care, precision agriculture, home automation, and military operations.
Since the number of sensor nodes in a WSN is in the order of hundreds or
thousands, such networks are generally considered fairly large.
Compared to other networks, WSNs operate on more limited communication,
computation and energy resources. Due to these limitations, planning energy
resources for nodes becomes one of the most critical issues in designing WSNs.
These networks are often working on small non-rechargeable batteries and are
expected to last for long periods of time, usually over several months or years.
They commonly organize themselves and transfer data through a shared medium using multi-hop communication.

Wireless systems such as Wi-Fi and Bluetooth share unlicensed spectrum with many WSNs, creating potential interference and contention. Cognitive radio (CR) allows a wireless device to observe its environment and adapt transmission parameters to use available spectrum\cite{akyildiz2006next}.
Cognitive technology provides access to new spectrum with better propagation
characteristics such as frequencies that are licensed for television broadcast, public services,
and so forth. Adapting parameters such as modulation type and constellation size can also improve communication \cite{5444432}.

Through a CR network, unlicensed spectrum users or secondary users (SUs)
periodically monitor the spectrum for white space and opportunistically
use the licensed wireless spectrum, which primarily is owned by the
licensed users or primary users (PUs). It is of utmost importance
that the operations of the secondary users do not hinder the operations of
the primary users. Routing must account for spectrum availability along a path, in addition to conventional measures such as hop count and delay. Route selection and spectrum selection are therefore coupled, and the routing protocol must account for PU activity.

It is well known that the one-to-many communication, i.e. multicast
routing, has important applications in WSNs. The multicast operation
in WSNs is usually just a sensor sending messages to multiple sensors
along a multicast tree. The source node sends messages to multiple
sinks, or a sink node sends messages to a set of sensor nodes. Code
update, assignment of tasks, and object query are some instances. Due to
the limited energy resources in WSNs, multiple unicasting can be replaced
with multicasting to improve network performance. Using multicast communication
results in optimization of the bandwidth consumption, less overhead and
energy conservation in the nodes \cite{Sarkar20163173,4052575}.

In order to address the aforementioned issues, we propose cognitive
multicast (CCast) as a novel multicast routing protocol in cognitive
wireless sensor networks (CWSN). CCast uses a latency-oriented spectrum-selection metric while seeking the energy benefits of shared multicast forwarding. This protocol is able to use
multicasting feature in CWSNs. It uses the virtual Euclidean Steiner-tree construction described in \cite{Wu20072931}.
The tree is constructed from the current root and destination coordinates, and routing uses local neighbor and spectrum information. A complete wireless connectivity map is not required.
This algorithm, like the routing protocol for cognitive radio ad hoc networks (CRP) \cite{5738222}, uses a single common control
channel (CCC) during route formation, while SUs are responsible for mapping
the characteristics of a delay function on the CCC.
Using such an approach, the overhead of route creation could be reduced.

In addition to CCast, we propose PCast which focuses on
minimizing power consumption. By concentrating on some factors such as
power consumption and channel conditions, PCast favors routes with lower normalized transmission and reception costs. Similar to CCast, this
protocol uses a common control channel (CCC) and a delay function.

Both proposed algorithms, i.e. CCast and PCast, are evaluated through MATLAB
simulations. The simulations compare energy consumption, total forwarding count, per-destination hop count, and packet-delivery delay.

The main contributions are the following. First, CCast combines geographic multicast grouping with the spectrum-aware RREQ forwarding-delay mechanism of CRP~\cite{5738222}. Second, PCast uses a normalized link-power cost to favor energy-efficient spectrum choices and routes. Third, both protocols use the rrSTR heuristic~\cite{1648836,Wu20072931} to group destinations, with local tree reconstruction at intermediate destinations. The rrSTR heuristic and the control-channel delay mechanism come from prior work. Our contribution is combining them for cognitive multicast and adding the PCast energy criterion. Finally, simulations show energy reductions of 21\% and 26\% relative to GMP for CCast and PCast, respectively, with longer routes and higher delay as the trade-off.

The rest of this paper is organized as follows: Section \ref{sec::relatedWork}
discusses related works while the design details for CCast and PCast protocols are described in Section \ref{sec::proposedAlgorithm}. The performance of these protocols is then evaluated
through simulation while the simulation details along with the result analysis are presented in Section \ref{sec::result}. Finally, the paper is concluded in Section \ref{sec::conclusion}.

\section{Related Work}
\label{sec::relatedWork}

In this section, we briefly review foundational CR routing protocols,
and further describe the traditional multicast protocols for wireless ad hoc and
sensor networks, in separate parts.

\subsection{Cognitive Radio Routing Protocols}

Classical routing metrics alone do not capture the constraints of cognitive wireless networks. Routing
protocols in cognitive wireless networks have to consider many issues
including PU activities which affect channel availability. The other
instance is the route selection over several channels such that PU operations
do not degrade. While there are many other instances, a lot of protocols are
proposed to face such issues. Cheng et al. \cite{4221542} proposed an on-demand
protocol for routing and spectrum assignment in cognitive radio networks. This protocol
implements on-demand routing and frequency band selection which satisfies the fact
that opportunistic spectrum access varies from node to node. Furthermore, the authors
of \cite{4221542} proposed a multi-flow, multi-frequency scheduling scheme and
defined cumulative delay along the path.

STOD-RP \cite{4698367} is then proposed to solve the challenge of collaboration
between the route selection and spectrum decision. Tree-based proactive routing
and on-demand route discovery are constructed by STOD-RP. This protocol reduces
the control overhead and the average of end-to-end delay.
Khalife et al. \cite{4698706} presented a routing approach for cognitive radio
networks based on probabilistic estimation. Routing in this protocol is designed
to improve route stability and availability. SEARCH \cite{Chowdhury20091983} is
further proposed as a completely distributed routing protocol for mobile CR networks.
This approach minimizes end-to-end path latency by cooperatively optimizing the path
and channel selections as well as considering the PU activity and mobility of the SU users.

Chowdhury and Akyildiz~\cite{5738222} proposed CRP for multi-hop cognitive-radio ad hoc networks. It accounts for PU protection and spectrum availability when selecting a route. Spectrum-selection metrics are mapped to RREQ forwarding delays over a common control channel, so that local conditions influence the arrival and accumulated cost of route requests. CCast adopts this mechanism.

Jiang et al. \cite{Jiang2016} proposed a clustering-based collaborative cognitive
routing algorithm (CCCR) capable of collaboratively
establishing the routes in multi-hop cognitive networks with several primary and
secondary users. Spachos and Hatzinakos~\cite{6750710} introduce a cognitive networking protocol with opportunistic routing for WSNs.
Related work on multicast protocols in multi-hop wireless networks is briefly
presented in the next part of this section.

\subsection{Multicast Protocols}

Forwarding strategy based multicast protocols are basically divided into two
categories, tree-based and mesh-based.
Kunz and Cheng~\cite{1022287} show that mesh-based protocols with additional
paths to forward multicast data packets provide high robustness at the cost of
additional transmission overhead and network loads. Works of
\cite{Lee603300, Garcia-Luna-Aceves:2006:CMP:2312104.2313040,
Chiang1998, 1392169} are examples of mesh-based protocols.
On the other hand, tree-based protocols construct a tree, rooted from the
source node and spanning to all destination nodes.
Such multicast protocols are well known to provide better data
forwarding efficiency at the cost of lower robustness in comparison with
mesh-based protocols. Hence, regarding the energy consumption, which is
considered as the major design objective in WSNs \cite{Sarkar20163173},
prominent protocols in multicast routing are considered the tree based ones.

Several tree-based protocols provide relevant building blocks. Position based
multicast routing (PBM) protocol proposed by Mauve et al. \cite{Mauve} is an instance.
In this algorithm the multicast tree is built based on a parameter denoted
as $\lambda$, which describes the trade-off between remaining overall distance
towards the destinations and bandwidth usage. Branch aggregation multicast (BAM)
\cite{1542820} is another tree-based routing protocol while its main contribution is in
the reduction of radio communication load. BAM combines single-hop aggregation and multiple-path aggregation to reduce communication load.
MgCast proposed by Mizumoto et al. \cite{1381901} is another tree-based
multicast algorithm. Given a source node and a set of geographical destination
regions, MgCast constructs a multicast routing tree that demonstrates the
trade-off between the number of route request messages and the number of links.

Geographic multicast routing protocol (GMP) which is introduced by Wu and Candan~\cite{1648836}
forms the multicast tree based on a virtual Euclidean Steiner tree. This protocol
uses a reduction ratio heuristic while, based on the transmitting nodes in this virtual
tree, destinations are divided into a set of groups. Regarding these groups,
packet forwarding nodes in the actual tree are specified. By taking the advantage
of multicasting in wireless networks, GMR \cite{4068104} improves the forwarding
efficiency. It also optimizes the cost over progress ratio, in which the cost is equal to
the number of selected neighbors and progress is the overall reduction of the remaining
distances toward destinations. In uCast which is proposed by Cao et al. \cite{4052575},
a unified connectionless multicast protocol for sensor networks is presented. This protocol
is robust, efficient and scalable against the topology changes.

Energy balanced multicast
routing (EBMR) which is presented by Wang et al. \cite{5662723}, is a WSN routing
algorithm with the goal of improving the energy efficiency and extending the network
lifetime for multicast forwarding. In EBMR, the node residual energy is used as a multicast forwarder-selection metric to construct the multicast tree.
Jiang et al. \cite{Jiang2015152} proposed network coding based multicast (NCM)
which employs network coding to improve energy efficiency of wireless networks.
This energy-efficient multicast routing algorithm increases network throughput and
reduces energy consumption with adaptable utilization of network structure.
A genetic algorithm is further used in \cite{6362126}
to consider energy consumption as well as end-to-end delay in mobile wireless networks
multicast routing. This algorithm applies only to source-based routing trees.

Later work also examines multicast in cognitive-radio networks. Phaswana et al.~\cite{phaswana2024} study a centralized multicast AODV approach for intermittent cognitive-radio ad hoc networks, using analytical models and simulations to compare AODV with multicast AODV (MAODV). Their setting differs from the geographic tree construction considered here.

\section{Proposed Algorithms}
\label{sec::proposedAlgorithm}

As we mentioned earlier, in this section we propose two new routing protocols
i.e. CCast and PCast. The protocols use local neighbor and spectrum information together with destination coordinates. Each node further is assumed to be aware of its location,
using a GPS device or instead the virtual coordinate assignment protocol defined
in \cite{1497887}. Furthermore, each source node in the multicast process knows the
location of all its neighbors as well as destination nodes, and PUs. On the other hand,
SUs are periodically sensing the environment for PU activities. They determine
the available channels that could be used for transmission. It is worthy to note that,
we use a common control channel (CCC) for exchanging various control data. Such data
includes spectrum sensing, neighbor discovery, maintenance information, and also
other information required for routing in CWSN.
This separates route discovery on the CCC from data transmission on the selected bands.
In both proposed algorithms, a Euclidean Steiner tree is used for constructing the multicast
tree. Computing an optimal Euclidean Steiner tree is NP-hard~\cite{garey1977}, thus many heuristic algorithms \cite{Zachariasen1999, Zelikovsky:1996:BAB:900518, Robins00improvedsteiner} are proposed to
approximately find the result of the problem. In this work, we use the rrSTR
algorithm proposed in \cite{1648836} to construct a heuristic
Euclidean Steiner tree.

After constructing the tree, we propose to identify some nodes as intermediate
destinations.
We use the identified intermediate destinations along with the final destinations to
send the packets. When the packets are received by intermediate destinations,
the multicast tree is constructed again. The new multicast tree contains also intermediate
destinations. This action basically is repeated and in each step the multicast tree
is refined according to the local knowledge of intermediate destinations. Meanwhile,
each intermediate node, i.e. one which is located between the source and the intermediate
destination, between an intermediate destination and another intermediate destination,
or between an intermediate destination and a final destination, uses a spectrum-selection
score to rank candidate forwarding nodes. This process imposes some delay on the end-to-end routing process. However,
accepting such a delay leads to reduced computational complexity at the
destinations.

Accordingly, the optimization function, which is required for the spectrum selection process,
has several constraints that should be satisfied. These constraints consist of
several metrics, which are proposed in \cite{5738222} and described in the rest of
this paper.
The first multicast routing algorithm, CCast, uses a latency-oriented spectrum-selection score for routing in CWSNs. The second one is PCast,
while its objective is minimizing energy consumption for multicast routing
in CWSNs. In the rest of this section, we briefly describe the rrSTR algorithm,
which is used regularly by the proposed protocols, in Part \ref{CMT}.
CCast and PCast protocols are also presented separately in Parts \ref{OCCast}
and \ref{OPCast}, respectively.

\subsection{Construct Multicast Tree}
\label{CMT}

Computing an optimal Euclidean Steiner tree is NP-hard~\cite{garey1977}. The virtual geometric tree uses terminal coordinates. Its branches are forwarded through the wireless network using local neighbor information. To solve this problem,
we use the rrSTR algorithm that is proposed in \cite{1648836}. The constructed tree
based on this algorithm only needs its own location beside its destinations' locations.
This approach is much cheaper than optimal calculations and is appropriate for geographic multicasting. The rrSTR heuristic repeatedly solves the three-terminal junction problem~\cite{hwang1992steiner}. For a triangle with an angle of at least $120^\circ$, the minimum-length junction is the corresponding terminal. Otherwise, it is an interior Steiner point. The tree may include vertices at which no actual wireless sensor nodes exist. So the multicast tree is virtual. In cognitive routing we utilize intermediate destinations (which may be virtual or actual nodes). The direct descendants of the root in the tree are intermediate destinations. In intermediate destinations, the Steiner tree is reconstructed based on its own position. This operation will have the opportunity to readjust the multicast tree. As we mentioned before, the rrSTR algorithm is used in our two proposed protocols. These protocols forward the packets after determining the intermediate destinations. Addressing the ultimate goal, the forwarding takes place differently in the protocols. After receiving packets in intermediate destinations, if another destination exists, the rrSTR algorithm reconstructs the multicast tree and intermediate destinations.

In this section we briefly describe the rrSTR algorithm. By using the reduction ratio, this algorithm creates the desired tree. The reduction ratio is based on the two following observations:
\begin{enumerate}
\item
It is possible for two arbitrary destinations to be close to each other, although sharing sub paths that are far from the source node.
\item
For two arbitrary destinations, it is possible to have a small angle between line segments connecting the source node to the destinations.
\end{enumerate}
Given a source node s and a set of two destinations (u,v), the reduction ratio, RR(s, u, v), is defined as follows:

\begin{equation}\label{eq:1}
RR(s, u, v) =  1 - \frac {d(s,t) + d(t,u) + d(t,v)}{d(s,u)  + d(s,v)}
\end{equation}
Here, $t$ is a minimum-length three-terminal junction, including the degenerate terminal case, and $d(x,y)$ is Euclidean distance. When the denominator is positive, $0\leq RR(s,u,v)\leq 1/2$. A larger reduction ratio indicates a greater relative geometric saving from sharing a branch for the selected pair. Based on the reduction ratio measure, the rrSTR algorithm constructs a Euclidean Steiner tree iteratively. In the first step source node s marks all destinations as active. In each iteration, a destination set with the largest RR value is selected. Then this set of destinations is marked as inactive, and the Steiner point of the nodes is added to the list of destinations and this node is marked as active. Also the corresponding two edges (the edges which include the set of destinations and the Steiner point of the nodes) are added to the tree. In the continuation of the algorithm these inactive destinations will not be considered, and the Steiner point that is added to the list of destinations is considered instead. This iteration continues as long as all destinations are covered. Further details of the heuristic are given in \cite{1648836}.

\subsection{Overview of CCast}
\label{OCCast}

Routing in CWSN compared to traditional WSN is facing several new challenges. Neighbor node detection is one of these challenges in cognitive networks. This challenge is due to the PUs' activities and dynamic changes of the environment, which result in a different active wireless system for each node. This difference among nodes in the network imposes a great overhead in each channel to cover all active wireless systems transmitting route requests (RREQs).
\begin{figure}[t]
\centering
\includegraphics[width=\linewidth]{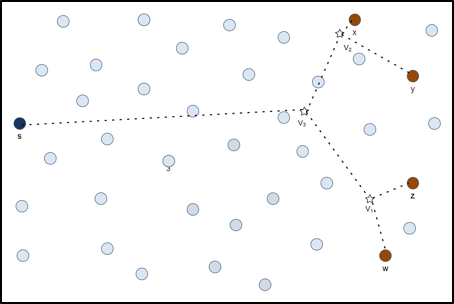}
\caption{Multicast tree constructed at the source node. Node s is a source node, nodes x, y, z, w are destinations and $v_3$ is an intermediate destination.}\label{fig:1}
\end{figure}

\begin{figure*}[t]
 \includegraphics[width=\textwidth]{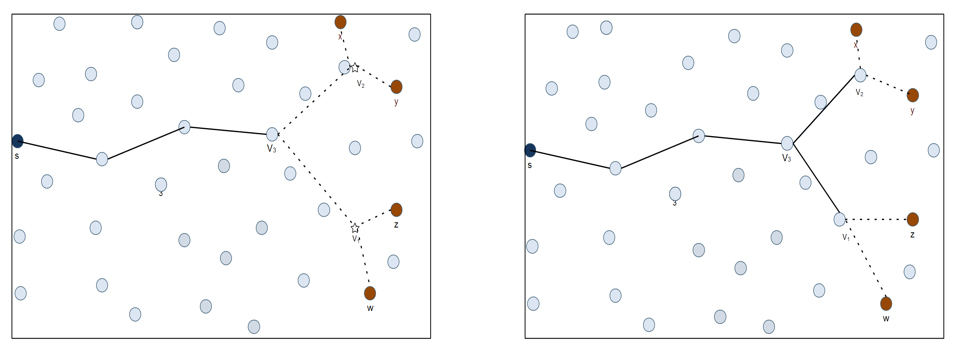}
\caption{(a) Node $v_3$ is assumed to be an intermediate destination and constructs a multicast tree based on the destination list. $v_2$ and $v_1$ are intermediate destinations. (b) Nodes $v_2$ and $v_1$ are assumed intermediate destinations and each construct multicast trees.}\label{fig:2}
\end{figure*}

Thus selecting the route needs spectrum information about other nodes participating in the routing. To solve this problem, a common control channel (CCC) can be used. Using the CCC enables exchanging various control data between source and destination, such as spectrum sensing, neighbor discovery, maintenance information and other information required for routing in CWSNs. The tree is constructed based on this routing information exchange. In this work, we use the CCC like \cite{5738222} to propagate the RREQ and RREP packets. The properties of the chosen spectrum that are independent of the number of possible spectrum bands are mapped to a delay function on the CCC. Also the neighbor nodes are discovered. This approach avoids sending a separate route request on every candidate data channel. Initially the source node, based on its location and destinations' locations, constructs the multicast tree, which is described in \cite{1648836}. Here we consider direct descendants of the root in the tree as intermediate destinations. These intermediate destinations refine the multicast tree and enable applying cognitive routing. The source node forwards a copy of the packet containing locations of the corresponding group of destinations' nodes to intermediate destinations. If a physical intermediate node is also a final destination, it removes itself from the remaining destination list after receiving the packet. If the list of destinations is not empty, this node based on its location and destinations' location constructs the tree and repeats the procedure described above. Figure \ref{fig:1} shows an example of a multicast tree that is constructed in the source node. In this figure, node s is a source node, nodes x, y, z, w are destinations and $v_3$ is an intermediate destination. At first, the source selects the pair of z and w then determines $v_1$ as the Steiner point. Then $v_2$ is selected as the Steiner point of the pair of x and y. At last, $v_3$ is selected as the Steiner point of the pair of $v_2$ and $v_1$. Figure~\ref{fig:2} shows the subsequent reconstruction: $v_3$ rebuilds the tree from its own position, and $v_1$ and $v_2$ in turn act as intermediate destinations for their groups. In order to forward a packet from the source node to each of the intermediate destinations, we use a similar approach to \cite{5738222}. This algorithm consists of two phases, which are as follows:
\begin{enumerate}
\item{Join query phase}
\item{Join reply phase}
\end{enumerate}
In the join query phase, an RREQ packet over the CCC is sent. The source node for each possible intermediate destination separately broadcasts the RREQ, which includes the corresponding list of destination nodes and other needed information. In each intermediate node, we use an optimization function that ranks candidate forwarding nodes by assigning a holding delay. Each intermediate node that receives this RREQ finds the best spectrum band based on local environmental observations and other aspects. Then an optimization function starts to calculate the value that the node should hold in a waiting state (DelayValue). When the DelayValue timer expires, the intermediate node broadcasts this packet. The optimization function and delay time are two factors that determine the waiting time of each node and establish a ranking of the route. The first RREQ favors a route with a small accumulated forwarding delay. To specify the total DelayValue of a path in an intermediate destination, the total DelayValue is added to the RREQ. In each intermediate node the DelayValue is added to the total DelayValue. We use the optimization functions which are proposed in \cite{5738222}. The objective of these optimization functions is to minimize the end-to-end CR route latency. The spectrum selection that uses these optimization functions has several constraints, which should be satisfied. These constraints consist of a number of metrics that are described in the rest of the paper.

This optimization function is defined as follows:
\begin{equation}
\label{eq:2}
OP = D_k \cdot T_f^x
\end{equation}

The objective is to maximize the optimization function that is described above. This score favors links with long range and long usable transmission time. In this equation $D_k$ is the propagation distance and $T_f^x$ is a fractional time for transmission that considers the different sensing schedules of the neighboring users.
A larger communication range can allow a route with fewer hops, depending on node locations and spectrum availability. Another metric that is used above is $T_f^x$. A larger value of this metric specifies a more effective duration in which transmission is allowed. This metric is calculated by the equation below:
\begin{equation}\label{eq:3}
T_f^x = \frac{T_a^x}{\max_{j\in\mathcal I_x\cup\{x\}}\{T_s^j + T_t^j\}}
\end{equation}
Here, $T_a^x$ is the effective transmission time available to SU $x$ on the selected band, taking neighboring sensing schedules into account as in CRP~\cite{5738222}. The set $\mathcal I_x$ contains the users whose schedules constrain $x$, and $T_s^j$ and $T_t^j$ are their sensing and transmission-window durations.

The spectrum k should satisfy the below equations:
\begin{equation} \label{eq:4}
M_B^k > (p_{\min})^{|C^k|}
\end{equation}
\begin{equation}\label{eq:5}
V_B^k < J_{\max}
\end{equation}
\begin{equation}\label{eq:6}
B_c^k > \psi_k \cdot |C^k|
\end{equation}
\begin{equation}\label{eq:7}
T_\Delta^s + T_\Delta^c (1-M_B^k) < T_{th}
\end{equation}
In Eq.~\eqref{eq:4} $M_B^k$ is the spectrum bandwidth availability probability based on the availability of the chosen $|C^k|$ channels in spectrum k in the node that is a candidate to forward, and $p_{\min}$ is a specified per-channel availability threshold. The exponent is the number of selected channels.

Assuming independent channel-availability states, $M_B^k$ is calculated as
\begin{equation}\label{eq:8}
M_B^k = \prod_{i\in C^k} p_i^k
\end{equation}
In the equation above $C^k$ is the set of channels chosen by the candidate forwarding node to meet user bandwidth requirements. Also the $p_i^k$ is the probability of finding the $i^{th}$ available channel. $p_i^k$ is calculated as below:
\begin{equation}\label{eq:9}
p_i^k = \frac{1/\beta_i}{1/\alpha_i+1/\beta_i}
      = \frac{\alpha_i}{\alpha_i+\beta_i}
\end{equation}
Here, $1/\alpha_i$ and $1/\beta_i$ are the mean PU on and off times for channel $i$ in band $k$. During an on interval the channel is occupied by a PU. During an off interval it is available to SUs. The availability probability is therefore the mean off time divided by the mean cycle duration.

In Eq.~\eqref{eq:5}, $J_{\max}$ is the largest permitted value of the bandwidth variance $V_B^k$. The value $V_B^k$ is calculated as below:
\begin{equation}\label{eq:10}
V_B^k = \psi_k  \sum_ {i \in C^k} \xi_i^k
\end{equation}
In the equation above, $\xi_i^k$ is the summation of the squares of the difference between the statistically known mean \textit{off} time and its currently observed off time, considering the past $N_v$ samples, for the $i^{th}$ channel in spectrum $k$.

In Eq.~\eqref{eq:6} the physical channel fading conditions are checked. $B_c^k$ is the coherence bandwidth and $\psi_k$ is the bandwidth of each channel in spectrum k. This condition limits frequency-selective fading.

Eq.~\eqref{eq:7} considers the switching time latency and checks this time to not be more than the predefined threshold $T_{th}$. $T_\Delta^s$ and $T_\Delta^c$ are the inter-band and intra-band switching delays, respectively.

As mentioned above, the source node constructs the tree and based on it, intermediate destinations are determined (direct descendants of the root in the tree are intermediate destinations). Regarding the intermediate destinations, the root splits destinations into some groups and creates a related RREQ packet for each group. Then it broadcasts the RREQ in the CCC for these intermediate destinations. Note that these intermediate destinations are constructed using the rrSTR algorithm and due to the lack of global knowledge these intermediate destinations may not exist in the real network. In other words, at the location of these virtual nodes there may not exist any actual wireless sensor nodes. To solve this problem, each intermediate node receiving an RREQ packet checks the distance between its location and the intermediate destination and, with the knowledge of neighbors' locations, calculates the distance between neighbors and the intermediate destination. If there exists a neighbor whose distance to the intermediate destination is less than the distance from the mentioned intermediate node to the intermediate destination, the procedure above continues until no such neighbor exists and the current node assumes the virtual intermediate-destination role.

Each intermediate node that receives an RREQ calculates the delay value based on the optimization functions. This delay value is then mapped into a delay scale that is divided into discrete segments. A segment indicates a constant forwarding delay that the corresponding node should wait for. Higher CCast spectrum-selection scores are mapped to shorter holding delays. This delay determines the ranking of a route. That is due to the fact that an RREQ in a higher quality route is received faster in the destination.

The nodes which received an RREQ and assumed themselves intermediate destinations check the list of destinations.

If the node is on this list, it removes itself from the list. Additionally, if the list of destinations is empty, the node becomes the last node in the multicast tree. Otherwise, the node assumes an intermediate destination based on its location and destinations location in the tree.

This node sends an RREQ to the linked intermediate destinations after constructing the tree. It also checks the total delay value of other RREQs that are received later against the total delay value of the first received RREQ. If there is an RREQ with a lower total delay value, it selects a new route. This occurs due to the MAC layer link's delays on the CCC.

When the last node removes itself from the destinations' list in the multicast tree, it sends an RREP packet based on the route that receives the RREQ. The intermediate destination that sends the RREQ, after receiving this RREP packet, establishes a route for this segment of the tree. If this destination sends another RREQ, it waits to receive the corresponding RREP packet. After receiving the RREPs for all required downstream branches, it sends an RREP to the preceding intermediate destination or source node.

In a sparse network, it is possible that several nodes become intermediate destinations. In this case, we apply a wait threshold to the source. During this time, multiple RREQs may arrive due to different routing paths. Based on the total delay value, the best route is selected. In a dense network, however, only the one node that is presumed intermediate destination does not need to wait for route selection. The first RREP establishes one branch. The source considers the complete multicast tree established after receiving replies for all required branches.
\subsection{PCast}\label{OPCast}
Energy consumption is the most important challenge in WSN. As such, we propose a power-aware multicast routing protocol in CWSN called PCast.
The objective for this cognitive multicast routing protocol is to reduce energy consumption of transmission in the path. Similar to CCast, in this protocol we use the CCC to overcome the problem in cognitive routing. As mentioned above, neighbor discovery, various control data, and routing information between source and destinations are exchanged over the CCC. This approach reduces the overheads of data required to create a multicast tree.
One of the benefits of cognitive radio is the possibility of using a lower frequency. Doing so, we can transmit at a lower power level and to further distances. Assuming the simple path-loss propagation model, we can use the following formula to calculate the transmission power:
\begin{equation}\label{eq:11}
P_{TX}=P_{RX} \Big(\frac{4\pi d f_k}{c}\Big)^2
\end{equation}
 where $P_{TX}$ is the transmit power under a free-space model with unit antenna gains, $P_{RX}$ is receiver sensitivity, $c$ is the speed of light, $f_k$ is the selected frequency, and $d$ is the link distance.
 Energy consumption rises with distance as long as frequency remains fixed. Similarly, for a certain distance, a higher frequency takes more power. Thus, lower frequencies can reduce the radiated transmit power for a fixed link distance. A complete route must also account for its number of transmissions and reception costs.

 PCast uses frequency-dependent link costs, PU activity, and the number of hops to rank candidate routes.
 In CCast, route construction uses a latency-oriented spectrum-selection score. Based on PUs' activity, this routing protocol uses a different frequency channel. In some cases, a route different from the CCast route may exist. This route goes around the area where PUs are activated and is allowed to use a lower frequency. Its lower per-link cost can outweigh the extra hops.

\begin{figure}[!ht]
\centering
\includegraphics[width=\linewidth]{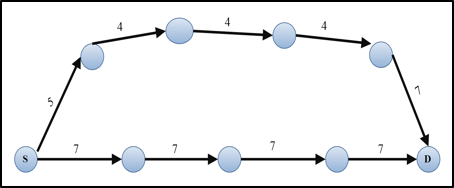}
\caption{Link costs for two routes from $S$ to $D$. The longer upper route has a lower total cost (24) than the lower route (28).} \label{fig:3}
\end{figure}

 This concept is presented in Figure \ref{fig:3} with two paths for forwarding S to D. The bottom path is a path where CCast is probably selected. Its illustrative total link cost is 28. The edge weights represent modeled communication costs in common units. The upper path is a path where PUs are not activated in a lower frequency. While this path is longer than the bottom path, due to using a lower frequency, the total illustrative cost drops to 24.
As described in the previous subsection, CCast uses optimization functions whose value is mapped to a delay scale. These delays determine the quality of a route and an early RREQ favors a route with a lower accumulated assigned delay. Similarly, in PCast we implement mapped optimization functions. The optimization functions used in this protocol, however, prefer the route with a lower frequency.

The total power consumption along a path includes the power consumption for transmitting and that for receiving. The power consumption for transmitting is based on the frequency and the propagated distance.

 The power consumption for receiving, however, is constant and independent of the applied frequency and distance. In this work, we use total power consumption as a metric to determine the quality of a route. The link-cost comparison assumes a fixed propagation distance. Under this assumption, a lower frequency requires less radiated transmit power. For equal packet airtime $\tau$, link energy is $\tau$ times the modeled transmitter-plus-receiver power.

 We use the following equation as the optimization function:
\begin{equation}
 f_r = \frac{P_c}{P_{\mathrm{ref}}}
 \end{equation}

 where $P_c$ is the modeled transmitter-plus-receiver power at the selected frequency, and $P_{\mathrm{ref}}>0$ is the corresponding power at the base frequency and fixed reference propagation distance. The reference power is common to the link-cost comparison under this fixed-distance model. The base frequency is the lowest frequency in the modeled band set. As such, the radiated transmit power for a given link distance is smallest at this frequency under the assumed propagation model.

 The objective for optimization is to find the spectrum k out of all possible spectrum bands that minimizes the optimization function. Similar to CCast, this k must satisfy Eq.~\eqref{eq:4} through Eq.~\eqref{eq:7}.

We define a shortest path problem on the weighted graph model of the network to find the path between a source and a destination so that the total weight of the path is minimized. PCast uses this additive-cost view to favor paths with a smaller summed modeled link cost. This problem is similar to the discussed graph problem where each node is a vertex of the graph and the power consumption between nodes is the weight of edges. To rank candidate paths, each relay adds a holding delay determined by its modeled link cost. If the weight of an edge is high we wait more time and vice versa.

 After finding the spectrum k that minimizes the optimization function, each node is weighted based on the delay determined as the value of the optimization function. The value of this delay, $D_e$, at each node is calculated as:
 \begin{equation}\label{eq:12}
 D_e = D_c \cdot f_r
 \end{equation}
 where $D_c$ is a constant delay.
 For example, in Figure~\ref{fig:3}, let the common reference cost be 4 and $D_c=0.1\,\mathrm{s}$. Link costs 4, 5, and 7 then give holding delays of 0.1, 0.125, and 0.175 s. Ignoring transmission, processing, and contention delays, the accumulated holding delay is $0.1(24/4)=0.6\,\mathrm{s}$ for the upper path and $0.1(28/4)=0.7\,\mathrm{s}$ for the lower path.

 The steps for constructing a multicast tree are similar to CCast, with the difference in forwarding the RREQ from a source node to an intermediate destination, where each intermediate node uses a delay value corresponding to the optimization function described above. Because all links share the same reference power, the accumulated delay is proportional to the total link power along the path.

\begin{figure}[t]
\centering
\includegraphics[width=\linewidth]{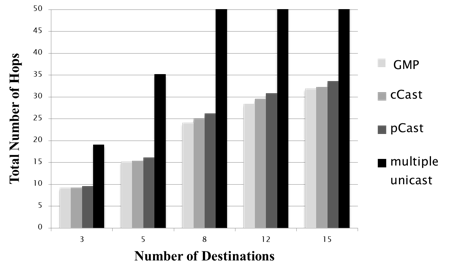}
\caption{Total number of hops in the multicast tree. GMP, CCast, PCast, and multiple unicast are compared. Some multiple-unicast bars exceed the plotted range of 50 hops. Their exact values are not shown. In Figures 4--7, the legend labels cCast and pCast denote CCast and PCast.} \label{fig:4}
\end{figure}

\section{Performance Evaluation}
\label{sec::result}
In this section, we evaluate whether the two proposed protocols utilizing cognitive radio consume less energy in simulation. Our simulation compares CCast and PCast with GMP and multiple unicast. GMP is the distributed geographic multicast routing protocol~\cite{1648836}. GMP uses the rrSTR algorithm to construct a multicast tree and in each hop repeats the rrSTR algorithm to construct the multicast tree. Multiple unicast sends a separate copy from the source to each destination. Multiple unicast uses \cite{Jiang2016} to route and reach each destination. Simulation results show that multicast routing consumes less energy than multiple unicast in CWSN and cognitive routing consumes less energy compared with traditional routing.

\subsection{Simulation Setup}
The MATLAB model contains 1000 SUs uniformly distributed over a $1000\,\mathrm{m}\times1000\,\mathrm{m}$ area. Each SU has a transmission range of 150 m, and the source and destination nodes are selected randomly. The evaluated destination counts are 3, 5, 8, 12, and 15. The area is divided into a $3\times3$ grid of equal squares, with a PU at the center of each square. These PUs freely activate and SUs observe this activity and leave the corresponding spectrum (i.e., perfect PU transmitter detection at the physical layer is assumed). Moreover, the transmission range for each PU is 150 m. We use an IEEE 802.11b-style MAC at a data rate of $1\,\mathrm{Mbit/s}$ in the modeled multi-band network. There are three different spectrum bands, each with six channels. The spectrum-selection framework and sensing/switching model follow CRP~\cite{5738222}. $D_k$ denotes the communication range associated with the selected band. Also, the RREQ forwarding delay has a segment size of $0.01\,\mathrm{s}$ and a maximum limit of $0.05\,\mathrm{s}$. Each SU has a sensing interval $T_s^i=0.1\,\mathrm{s}$ and a transmission window $T_t^i=0.6\,\mathrm{s}$. The inter-band switching delay is $T_\Delta^s=1\,\mathrm{ms}$, the intra-band switching delay is $T_\Delta^c=200\,\mu\mathrm{s}$, and $T_{\mathrm{th}}=1.2\,\mathrm{ms}$. With these values, the switching constraint in Eq.~\eqref{eq:7} is satisfied whenever Eq.~\eqref{eq:4} holds.

\begin{figure}[t]
\centering
\includegraphics[width=\linewidth]{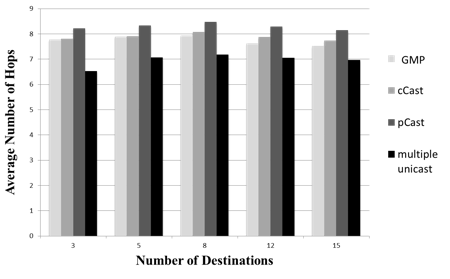}
\caption{Hop count per destination. Multiple unicast protocol is shown to require the least amount of hops for each destination.} \label{fig:5}
\end{figure}

\subsection{Total Number of Hops in the Multicast Tree}
The total number of hops is the number of necessary forwardings for a packet to arrive at all destinations. The four mentioned protocols (GMP, CCast, PCast and multiple unicast) are compared with regard to the total number of hops and the results are shown in Figure \ref{fig:4}.
The simulation shows that GMP needs a lower total number of hops to deliver a packet to all destinations. This result is due to GMP using the rrSTR algorithm and, in each intermediate node, refining the constructed multicast tree. The result for CCast is very close to that for GMP. Since CCast does not refine the multicast tree at each ordinary relay (the multicast tree is refined in intermediate destinations), in some cases an extra hop may be created in the multicast tree. Whenever possible, PCast goes around the area with active PUs and uses a lower frequency, and consequently finds the path that consumes lower energy. Thus, the multicast tree that is constructed during PCast requires more hops to reach all destinations. Multiple unicast finds routes independently for each destination and does not share the common paths, which reduces forwarding in the multicast routing. Therefore, as shown in Figure \ref{fig:4}, using the multiple unicast protocol increases the hop count.

\subsection{Per-Destination Hop Count}
\label{sec:per-destination}
The simulation results for the average number of hops for each destination are plotted in Figure \ref{fig:5}. In this plot, the multiple unicast protocol is shown to require the least amount of hops for each destination. This is because in this protocol a packet is delivered to its destinations independently and, unlike multicasting, does not travel along longer paths to save energy. Compared with CCast and PCast, GMP needs a lower number of hops to reach each destination. The hop count per destination for CCast is slightly higher than GMP. Similarly, the average number of hops for each destination for PCast is the highest due to its chance of selecting a far path.

\subsection{Energy Consumption}
\begin{figure}[t]
\centering
\includegraphics[width=\linewidth]{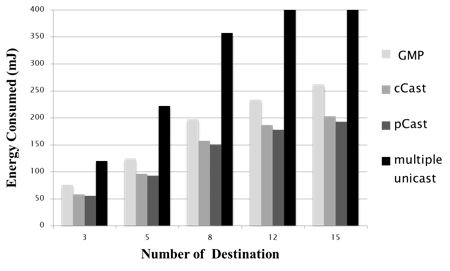}
\caption{Average modeled energy consumed during delivery to all destinations, in mJ. Some multiple-unicast bars exceed the plotted range of 400 mJ. Their exact values are not shown.} \label{fig:6}
\end{figure}

The metric for energy consumption is the average energy consumed during packet delivery to all destinations. The simulation results for energy consumption are shown in Figure \ref{fig:6}. The multiple unicast routing protocol consumes the most energy.
The number of hops represents the number of dispatches required to reach destinations. As such, due to its higher hop count, multiple unicasting consumes more energy than the other protocols. The total number of hops for CCast and GMP routing protocols are very close to each other. However, due to utilizing cognitive radio for multicasting, CCast consumes less energy than GMP. CCast uses 21\% less energy than GMP. PCast can avoid regions where PUs occupy a low-frequency band, allowing the route to use that band where it remains available. As it is specified in Figure \ref{fig:6}, PCast is consuming the least energy. The reduction in energy consumption for PCast compared to GMP is 26\%.

\subsection{Routing Delay}

\begin{figure}[t]
\centering
\includegraphics[width=\linewidth]{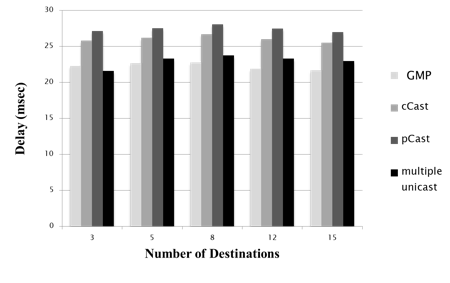}
\caption{Mean packet-delivery delay, in ms, for the simulation comparison. PCast has the largest delay among the evaluated protocols.} \label{fig:7}
\end{figure}
The packet-delivery delay is averaged over destination nodes. It does not include the RREQ holding delays used during route discovery. Based on the simulation results shown in Figure \ref{fig:7}, it is concluded that the average number of hops for each destination plays an important role in the routing delay. As Figure~\ref{fig:7} shows, multiple unicast has lower delivery delay than CCast and PCast. Despite its latency-oriented spectrum-selection score, CCast's delivery delay exceeds GMP's in these simulations. The cognitive-radio methods face sensing and switching constraints in addition to the effect of route length. Moreover, PCast has the highest routing delay consistent with its larger per-destination hop count in Section~\ref{sec:per-destination}.

Our evaluation has several limitations. The simulations assume perfect detection of PU transmitters and use a free-space link-power model, so they do not capture hardware energy consumption or interference at undetected PU receivers. The PCast link cost assumes a fixed link distance and selects routes segment by segment. It therefore does not guarantee a minimum-energy multicast tree. GMP does not use cognitive radio, so part of the energy difference between GMP and the proposed protocols comes from access to additional spectrum. Finally, the forwarding rule toward virtual intermediate destinations does not handle geographic voids.

\section{Conclusion}
\label{sec::conclusion}
In this paper, we proposed geographic multicast routing for cognitive wireless sensor networks called CCast. CCast ranks forwarding choices with a latency-oriented spectrum-selection score. Its energy savings come from shared multicast forwarding and access to additional spectrum. This protocol uses destination coordinates and local neighbor information to construct a virtual multicast tree and route its branches through physical nodes. In simulation, CCast used 21\% less energy than GMP. Furthermore, CCast with multicast forwarding consumes less energy than multiple unicasting in cognitive wireless sensor networks.
PCast was shown to save modeled communication energy by favoring lower-cost routes, including detours around regions where PUs occupy low-frequency bands. The performance evaluation for the PCast protocol showed that this protocol consumes 26\% less energy than GMP. It also consumed less energy than multiple unicast in the evaluated configurations. These gains come at the cost of longer per-destination paths and higher delivery delay.

\begingroup
\small
\bibliographystyle{ieeetr}
\bibliography{references_mc}
\endgroup
\end{document}